\documentstyle[prb,tighten,aps]{revtex}
\begin{document}
\draft
\title{Open rigid string with the Gauss-Bonnet term in action}
\author{
V.~V.~Nesterenko\thanks{Electronic address:
nestr@thsun1.jinr.dubna.su},
I.~G.~Pirozhenko\thanks{Electronic address:
pirozhen@thsun1.jinr.dubna.su}}
\address{Bogoliubov Laboratory of Theoretical
Physics, JINR \\  141980 Dubna Russia}
\date{\today}
\maketitle
\begin{abstract}
The effect of the Gaussian curvature in the rigid string action on
the interquark potential is investigated. The linearized equations of
motion and boundary conditions, following from the modified string
action, are obtained. The equation, defining the eigenfrequency
spectrum of the string oscillations is derived. On this basis, the
interquark potential, generated by the string is calculated in
one-loop approximation. A substantial influence of the topological
term in the string action on the interquark potential at the
distances of hadronic size order or less is revealed.
\end{abstract}


\section{Introduction}
The difficulties in the  Nambu--Goto string quantum theory, such as
the nonphysical space--time dimension and the tachyonic state  in the
string spectrum, are well known. The need for an adequate description
of the quark interaction in hadrons initiated the appearance of the
rigid string model.  This model was suggested by A.~M.~Polyakov and
independently by H.~Kleinert \cite{c1}. Due to its finite thickness,
the rigid string is characterized not only by its tension but also by
its resistance to transverse bending (rigidity). This is incomplete
analogy with classical dynamics of rods and beams. However the energy
of Polyakov--Kleinert rigid string proves to be  unbounded from
below~\cite{c2} because of the second derivatives in the string
action. Therefore only Euclidean rigid string model is well defined.

In the applications to hadronic physics open strings are to be
considered~\cite{c3}. Here an important role is played by boundary
conditions on the string dynamical variables (string coordinates).
For example, when the boundary terms describing point-like masses
on the string ends are added to the Nambu--Goto, the interquark
potential is considerably modified. In the case of extremely
asymmetric quark mass configuration this results in the removal
of the tachyonic state contribution to the string
potential~\cite{c4}.

It was specifically supposed that the boundary conditions in the
effective rigid sting model following from QCD enable one to suppress
the oscillation modes giving negative contribution to the
energy~\cite{c5}. Unfortunately, this idea has not been implemented
yet.

Another approach to the problem of energy unboundness from below was
considered in~\cite{c6}, where the rigid string model with nonlocal
action was put forward.

A consistent way to introduce the boundary conditions into the string
dynamics is to add the corresponding terms (geometrical invariants)
to the initial string action~\cite{c7,c8,c9}. In this case the
boundary conditions are consistent with the dynamical equations for
sure.

The first candidate to modify the boundary conditions in
string models is obviously the Gaussian curvature of string
world surface.  This geometrical invariant depends
on the second derivatives of the metric induced on the string
world sheet. As the rigid string action contains
the second derivatives6 it is natural to modify it with
the Gaussian curvature term. According to the Gauss--Bonnet
theorem~\cite{c10} the surface integral of the gaussian curvature
can be reduced to a contour integral along the boundary of the
surface. As a result, for closed  strings this term in the action
gives the Eulerian characteristics of the string world sheet.
For open strings it was shown that, when  the action is modified
in such a way, the string ends can not move with the velocity
of light, as they do in the Nambu--Goto model with free
ends~\cite{c11}.

A general mathematical analysis of the boundary conditions in the
rigid string model was performed in~\cite{c8}.  However, the
influence of these conditions on the concrete physical predictions is
not properly studied. Only particular results were obtained
here~\cite{c9},\cite{c9a}.

The aim of the present note is to investigate the effect of the
Gaussian  curvature in the rigid string action on the interquark
potential generated by the string. The layout of the paper is the
following. In Section 2 the linearized equations of motion and
boundary conditions in the rigid string model with the action
modified by the Gaussian curvature are derived. Further the equation
defining the eigenfrequencies of the string oscillations is obtained.
Proceeding from this equation, in Section 3 the one--loop interquark
potential, generated by the string is found. By making use of the
numerical calculations, it is shown that the modification of the
string action by the topological term considerably changes the
interquark potential at the distances comparable with the size of
hadrons or less. Section 4 is devoted to the discussion of the
results obtained.  \section{Modification of the boundary conditions
by the Gaussian curvature in the rigid string action} The action of
the relativistic string with rigidity is the following~\cite{c1}
\begin{equation}
S\,=\,- M^{2}_0 \int \int d^{2} \xi \sqrt{-g}\left[1
- \frac{\alpha}{2 M_0^2} \triangle x^{\mu}\triangle x_{\mu}\right].
\label{a1}
\end{equation}
Here $M^2_0$ is the string tension,
$\alpha$ is a dimensionless parameter characterizing the string
rigidity, $x^{\mu}(\xi_0,\xi_1)$ are the string coordinates in
$D$-dimensional space--time, $\mu=0,1,...D-1$.  The curvilinear
coordinates $\xi_0, \xi_1$ are introduced on the string world--sheet.
The imbedding of the string world surface into the enveloping
space--time induces a metric on this surface $g_{ij}(\xi)=\partial_i
x^{\mu} \partial_j x_{\mu},\;i,j=0,1;\;g_{ij}g^{jk}=\delta^k_i;\;
g=\det g_{ij}$. The Laplace--Beltrami operator related to the induced
metric is defined by
\begin{equation} {\triangle} \,= \, \frac{1}{
\sqrt{-g}} \frac{ \partial}{{ \partial} {\xi}^i} \left(\sqrt{-g} \;
g^{ij} \frac{\partial}{{\partial} {\xi}^i}\right).
\label{a2}
\end{equation}

Let us add to the action (\ref{a1}) a topological term, proportional
to the integral Gaussian (intrinsic) curvature  of the string world
surface
\begin{equation}
-\beta \int \int d^{2} \xi \sqrt{-g} \;K.
\label{a3}
\end{equation}
By the Gauss--Bonnet theorem(\ref{a3})
this term can be transformed into the integral along a closed
contour $\partial \Omega$ bounding the string world surface
\begin{equation} \int \limits_ \Omega^{} d^{2} \xi
\sqrt{-g} \; K \,= \,- \oint \limits_ {d \Omega}^{} k_g \;d \,s \;+\;
\mbox{const},
\label{a5}
\end{equation}
where $k_g$  is the geodesic curvature.
For a curve lying on a surface and defined by natural parametrization
${\bf r}(s), (d{\bf r}/d s)^2=1$, the  geodesic curvature, $k_g$, is
given by the formula~\cite{c10}
\begin{equation}
k_g^2=({\bf k}_{\|})^2,
\label{a6}
\end{equation}
where ${\bf k}_{||}$ is a tangential component of the curvature
vector
\begin{equation}
{\bf k}\equiv\frac{d^2{\bf r}}{d s^2}={\bf k}_{\bot}+{\bf k}_{\|}.
\label{a7}
\end{equation}

Now we choose the coordinate set on the string world surface in a
way that the trajectories of the string ends were defined by a
condition $\xi_1=\mbox{const}$. then the geodesic curvature of
these trajectories can be expressed through the components of
the metric tensor $g_{ij}$
\begin{equation}
k_g=-\frac{1}{2}\,\frac{g_{00}\acute{g}_{00}-2 g_{00}\dot{g}_{01}
+g_{01}g_{00}}{\sqrt{-g}(g_{00})^{3/2}}.
\label{a8}
\end{equation}

Further we shall use the nonparametric (Gaussian) definition
of the string world sheet
\begin{equation}
x^{\mu}(\xi_i)=(\xi_0,\xi_1,x_2,\dots x_{D-1})=
(\xi_i,{\bf u}(\xi_k)),
\label{a9}
\end{equation}
$$i=0,1,\;\;\xi_0=t,\;\; \xi_1=r,\;\;0<r<R.$$
In this parametrization the geodesic curvature is given by
\begin{equation}
k_g\;=\;-\; \frac{\ddot{\bf u} \;
\acute{\bf u}}
{\sqrt{1+{\acute{\bf u}}^{2}-{\dot {\bf u}}^{2}} \;
{(1-{\dot{\bf u}}^2)}^{3/2}},
\label{a10}
\end{equation}
where $\dot{\bf u}=\partial u/\partial t,\;\acute{\bf u}=
\partial u/\partial r$.

In what follows we shall treat $\dot{\bf u}$ ¨ $\acute{\bf u}$
as small quantities
\begin{eqnarray}
\sqrt{-g}& =& \sqrt{1-{\dot{\bf u}}^{2}+{\acute{\bf u}}^{2}} \simeq
 1- \frac{1}{2} {\dot{\bf u}}^{2}+\frac{1}{2} {\acute{\bf u}}^{2},
\nonumber\\
\frac{1}{\sqrt{-g}}&\simeq& 1+\frac{1}{2}
{\dot{\bf u}}^2-\frac{1}{2} {\acute{\bf u}}^2.
\label{a11}
\end{eqnarray}
Taking into account (\ref{a11}) the string action in harmonic
approximation acquires the form
\begin{equation}
S=-M_0^{2} \int \limits_0^{R}dr \int \limits_{t_1}^{t_2}dt\left[1+
\frac{1}{2} ({\acute{\bf u}}^2-{\dot {\bf u}}^2)+
\frac{\alpha}{2 M_0^2}
{(\Box {\bf u})}^2\right]- \beta \int
\limits_{t_1}^{t_2} dt \; \ddot{\bf u} \acute{\bf u} ,
\label{a12}
\end{equation}
where $\Box=\partial^2/\partial t^2-\partial^2/\partial r^2$ is
the two dimensional d'Alambert operator. The action (\ref{a12})
gives rise to the linear equations of motion
\begin{equation}
\Box (M_0^2+\alpha  \Box  ) {\bf u} =0
\label{a13}
\end{equation}
and the boundary conditions
\begin{eqnarray}
\frac{\partial}{\partial r}\left(M_0^2{\bf u}+\beta\ddot{\bf u}+
\alpha\Box{\bf u}\right)&=&0,\;\;\;r=0,R,
\label{a14}\\
\beta\ddot{\bf u}-\alpha\Box{\bf u}&=&0,\;\;\;r=0,R.
\label{a15}
\end{eqnarray}
Due to the second derivatives in the rigid string action the number
of obtained boundary conditions is twice compared with the
Nambu--Goto case.

The solutions of the boundary value problem  (\ref{a13}--\ref{a15})
can  be sought in the form
\begin{equation}
{\bf u}(r,t)\sim C e^{i\omega t+ikr}.
\label{a16}
\end{equation}
Substituting (\ref{a16}) into the equations of motion (\ref{a13}),
one obtains the dispersive equation
\begin{equation}
(\omega^2-k^2)[M_0^2-\alpha(\omega^2-k^2)]=0,
\label{a17}
\end{equation}
with four branches of solutions
\begin{eqnarray}
k_1=\omega,&\;\;&k_2=-\omega,
\label{a18}\\
k_3=\Omega,&\;\;&k_4=-\Omega,\;\;\;\;
\Omega=\sqrt{\omega^2-M_0^2/\alpha}.
\nonumber
\end{eqnarray}
Now the general solution to the string coordinates (\ref{a16})
can be rewritten as
\begin{equation}
{\bf u}(t,r)={\bf u}_0 e^{i\omega t}
\sum\limits_{j=1}^{4}C_j e^{i k_j r},
\label{a19}
\end{equation}
where ${\bf u}_0$ is a constant vector, and $C_j$ are amplitudes
determined by initial conditions.

The solutions of Eq.~(\ref{a13}) should satisfy the boundary
conditions (\ref{a14}) and (\ref{a15}). The substitution of
(\ref{a19}) into (\ref{a14}) and (\ref{a15}) results in a system of
linear homogeneous equations for the amplitudes $C_j$:
\begin{eqnarray}
&&
q C_1 -  q C_2 - \beta \omega \Omega C_3+\beta\omega\Omega C_4=0,
\nonumber\\
&&q e^{i\omega R}C_1 -  q e^{-i\omega R} C_2
- \beta\omega\Omega e^{i\Omega R} C_3+
\beta\omega\Omega e^{-i\Omega R} C_4=0,
\nonumber\\
&&\beta\omega^2 C_1+\beta\omega^2 C_2- q C_3-q C_4=0,
\nonumber\\
&&\beta\omega^2 e^{i\omega R}C_1+\beta\omega^2 e^{-i\omega R} C_2-
q e^{i \omega R}C_3-q e^{-i \omega R}C_4=0,
\label{a20}
\end{eqnarray}
where $q=M_0^2-\beta\omega^2$.
The equation for the eigenfrequencies is obtained by setting the
determinant of the system (\ref{a20}) equal to zero
\begin{eqnarray}
f(\omega)&\equiv&\sin(\omega R)\sin(\Omega R)[(M_0^2-
\beta \omega^2)^4+\beta^4\omega^6\Omega^2]\nonumber\\
&-&
2 (M_0^2-\beta\omega^2)^2\beta^2\omega^3\Omega
[1-\cos(\omega R)\cos(\Omega R)]=0.
\label{a21}
\end{eqnarray}
This equation is rather complicated. Nevertheless it contains
information that enables one to make concrete physical predictions in
the framework of the string model in hand.

\section{String potential} Making use of the string eigenfrequencies
we are able to calculate the first quantum correction to the energy
of the  string ground state, i.e. the Casimir energy of the system.
Considering this energy as a function of the string length $R$ we
arrive at the interquark potential generated by the string.  The
interquark potential $V(R)$ introduced in this way is defined by the
formula~\cite{c11}
\begin{equation}
\exp[-V(R)/T]=\int[{\cal D}{\bf u}]\exp\{-S^T[{\bf u}]\},\;\;T\to0,
\label{a22}
\end{equation}
where
\begin{equation}
S^T=M_0^2\int\limits_0^{1/T}dt\int\limits_0^R dr
\left[1+\frac{1}{2}{\bf u}\left(1-\frac{\alpha}{M_0^2}\triangle\right)
(-\triangle){\bf u}\right]-\beta\int\limits_0^{1/T}
dt\,\ddot{\bf u} \acute{\bf u}
\label{a23}
\end{equation}
is the Euclidean string action, $\triangle=\partial^2/\partial t^2+
\partial^2/\partial r^2$ is the two--dimensional Laplace
operator, $T$ is the temperature. The functional integration
is carried out  over the string coordinates obeying the
periodicity condition
${\bf u}(t,r)={\bf u}(t+1/T,r)$. The Euclidean equations of motion
and boundary conditions are obtained from (\ref{a13}), (\ref{a14}),
and (\ref{a15}) by means of substitution $t\to it$. It is easy to
demonstrate that the Euclidean eigenfrequencies are defined by the
same equation (\ref{a21}).

After functional integration in (\ref{a22}) we derive the expression
for the string potential
\begin{equation}
V(R)=M_0^2 R+\frac{D-2}{2\beta} \mbox{Tr}\ln\left[\left(1-
\frac{\alpha}{M_0^2}\triangle\right)(-\triangle)\right],
\;\;T\to 0.
\label{a24}
\end{equation}
The boundary term (\ref{a5}) in the action is taken into account
when calculating the eigenfrequencies of the operator
$(1-\alpha\triangle/M_0^2)(-\triangle)$, which determine the
functional trace in (\ref{a24}).

In the limit $T\to 0 $ we have~\cite{c3}
\begin{equation}
V(R)=M_0^2 R+\frac{D-2}{2}\sum\limits_{n=1}^{\infty}\omega_n,
\label{a25}
\end{equation}
where $\omega_n$  are the roots of the equation (\ref{a21}).
The first term in the formula (\ref{a25}) is the classical
string energy proportional to its length (confining potential),
The second term is the Casimir energy in the string model under
consideration.
The string potential given by Eq.(\ref{a25}) obviously requires
renormalization because the sum of the eigenfrequencies diverges.
The renormalized string potential at large distances should coincide
with its classical value
\begin{equation}
\left.V^{ren}(R)\right|_{R\to\infty}=M^2 R,
\label{a26}
\end{equation}
where $M^2$ is the renormalized string tension~\cite{Just}.  Starting
with Eq.~(\ref{a25}) and taking into account the necessity to
regularize all the divergent expressions, one arrives at the
following formula for the renormalized string  potential
\begin{eqnarray}
V^{ren}(R)&=&M_0^2 R+\left.(D-2)E_C^{reg}(R,
\Lambda)
\right|_{\Lambda\to\infty}\nonumber\\
&=&M_0^2 R+(D-2)\left\{E_C^{reg}(R,\Lambda)-
\left.E_C^{reg}(R\to\infty,\Lambda)\right\}\right|_{\Lambda\to\infty}
\nonumber\\
&&+\left.(D-2) E_C^{reg}(R\to\infty,\Lambda)\right|_{\Lambda\to\infty}
\nonumber\\
&=&M^2 R+(D-2)E_C^{ren},
\label{a27}
\end{eqnarray}
where $\Lambda$ is the regularization parameter,
$M^2$ is the renormalized value of the string tension
\begin{equation}
M^2=M_0^2+\frac{D-2}{R}\left.E_C^{reg}(R\to\infty,
\Lambda)\right|_{\Lambda\to\infty}.
\label{a28}
\end{equation}
The renormalized Casimir energy is defined by
\begin{equation}
E_C^{ren}(R)=\left.\left[E_C^{reg}(R,\Lambda)-
E_C^{reg}(R\to\infty\Lambda)\right]\right|_{\Lambda\to\infty},
\label{a29}
\end{equation}
where $M_0^2$ should be substituted by $M^2$ according to (\ref{a28}).
The sum (\ref{a25}) can be represented  in terms of the contour
integral by making use of the Cauchy theorem
(argument principle)~\cite{Lavr}
\begin{equation}
\frac{1}{2}\sum\limits_0^{\infty}\omega_n=
\frac{1}{4\pi i}\oint\limits_C^{}
d\omega\ln [f(\omega)]
\label{a30}
\end{equation}
where the function $f(\omega)$ is given by (\ref{a21}), and the
contour $C$ encloses the zeros of $f(\omega)$ situated in the right
half-plane of complex variable $\omega$. The frequency equation
(\ref{a21}) is real, therefore according to the Riemann--Schwarz
theorem~\cite{Lavr} its complex roots are lying symmetrically with
respect to the real axis. Consequently under the summation their
imaginary parts are mutually cancelled and the Casimir energy proves
to be real.

Because of the square root, the function (\ref{a21}) is obviously
two--valued. To select its single--valued branch, we make a cut
connecting the branch points
$\omega_0=\pm M/\sqrt{\alpha}$ along the real axis.
After that the contour can be chosen as it is shown in
Fig.~1. The radius of the contour $\Lambda$
stands for  a renormalization parameter for the divergent sum
$1/2\sum_{n=1}^{\infty}\omega_n$. Integration along the semicircle
in  $\Lambda\to\infty$ limit contributes only to the
counterterm~\cite{Just}.
The integrals along the edges of the cut are mutually cancelled,
and the integration around the branch point
$\omega=\omega_0$ gives the $R$--independent constant.
The integration along the interval
$(-i\Lambda,i\Lambda)$ produces the Casimir energy of the rigid
string with modified action
\begin{eqnarray}
\lefteqn{E_C^{reg}(M,\alpha,\beta,R)=\frac{1}{2\pi}
\int\limits_0^{\Lambda}dy
\ln\left(f(iy)\right),}
\label{a31}\\
f(iy)&=&-\sinh(R y)\sinh(R\bar\Omega)[(M^2+\beta y^2)^4
+\beta^4 y^6\bar\Omega^2]\nonumber\\
&&-2(M^2+\beta y^2)^2\beta^2 y^3\bar\Omega
[1-\cosh(Ry)\cosh(R\bar\Omega)],\nonumber
\end{eqnarray}
where $\bar\Omega=\sqrt{y^2+M^2/\alpha}$. For the renormalized
value of the  Casimir energy to be obtained, the asymptotics of
(\ref{a31}) when $R\to\infty$  is needed
\begin{equation}
E_C^{reg}(R\to\infty)=\frac{1}{2\pi}\int\limits_0^{\Lambda}
dy \ln\left\{-e^{Ry} e^{R\bar\Omega}\left[(M^2+\beta y^2)^2-
\beta^2 y^3 \bar\Omega\right]^2\right\}.
\label{a32}
\end{equation}
After the subtraction (\ref{a29}) we are left with the renormalized
Casimir energy
\begin{eqnarray}
E_C^{ren}(M,\alpha,\beta,R)
&=&\frac{1}{2\pi}
\int\limits_0^{\infty}
dy \ln \left\{ \left(1-e^{-2 R y}\right)
\left(1-e^{-2 R \bar \Omega}\right)\right.\nonumber\\
&&\;\;-\left.F(\alpha,\beta,M^2,y)
\left(e^{-Ry}-e^{-R \bar\Omega}\right)^2\right\},
\label{a33}
\end{eqnarray}
where
$$
 F(\alpha,\beta,M^2,y)=\frac{4 \beta^2 y^3\bar\Omega
(M^2+\beta y^2)^2}{[(M^2+\beta y^2)^2-\beta^2 y^3 \bar\Omega]^2}.
$$
This expression determines in the one-loop approximation
the first quantum correction to the classical linearly rising string
potential $V(R)\sim M^2 R$.
When $\beta=0$ the function $F$ vanishes, and Eq.~(\ref{a33})
reduces to the one--loop Casimir energy of rigid string ~\cite{c3}.
It is rather difficult to examine Eq.~(\ref{a33}) analytically
therefore we turn below to numerical calculations.
The interquark potential $V(\rho)/M,\;\rho=M R$ for different values
of the parameters $\alpha$ and $\beta$ (formulae
(\ref{a27}) and (\ref{a33}) with $D=4$) is presented in Fig.~2.
The potential generated by the Polyakov--Kleinert rigid string
$(\beta=0)$ for $\alpha=1,\,10,\,100$
is plotted in Fig.~2$a$. The values of the parameter
$\alpha$ are chosen with allowance for the following considerations.
In Abelian gauge model with simple  Higgs potential
(Nielsen--Olesen vortex model for relativistic string~\cite{Greg})
it was shown that the ratio $\alpha/(r_s^2 M^2)$, where $r_s$ --
is the gluonic tube radius, approximately equals $20$.
Keeping in mind that the quantity $M^{-1}$  gives the hadronic
size in string models, one can put
$r_s\sim(1/3)M^{-1}$. After that for the parameter
$\alpha$  we obtain $\alpha\sim 2$.

Figures 2$b,c,d$  show the impact of the Gaussian curvature term
in the string action on the interquark potential.
With increasing  $\beta$ the potential curves are shifted to the
right with respect to the interquark potential generated by
Polyakov--Kleinert string $(\beta=0)$.
For small $\beta $ it is easy to calculate the value of this shift
\begin{equation}
\delta V(R)=-\frac{2(D-2)\beta^2}{M^4}
\int\limits_0^{\infty}dy \,y^3 \bar\Omega
\frac{\left(e^{-R y}-e^{-R\bar\Omega}\right)^2}{
\left(1-e^{-2 R y}\right)\left(1-e^{-2R\bar\Omega}\right)}<0.
\label{a34}
\end{equation}

The potential curves in Figs. 2$b,c,d$ testify to obvious
correlation between the values of the parameters $\alpha$ and $\beta$.
To be exact, for fixed $\alpha$ one can obtain the same alteration
of the potential by setting $\alpha\sim\beta$.
Of course, this correlations appears only at the qualitative level.
The curves in Fig.~2 convincingly demonstrate that the modification
of the boundary conditions due to the Gaussian curvature in the string
action leeds to a considerable alteration of the interquark potential
at the distances $\le M^{-1}$.
In this range the effect of the Gaussian curvature term turns out
to be  comparable with the transition from the Nambu--Goto
model to the rigid string.
At large distances $R\to\infty$ all the potential curves
tend to the same asymptotics $\sim M^2 R$.

\section{Conclusion}
In view of an important role of the modification of the string
action in hand it is natural to inquire oneself about the physical
meaning of the parameter $\beta$. As this parameter enters only
the boundary conditions for string variables6 one can treat it
as a coupling constant characterizing the residual quark
interaction with gluonic field. This interaction has been ignored when
the collective string variables were introduced. Obviously it is
essential at small distances were localized gluonic tube (string)
does not reproduce adequately the real physical picture.
\acknowledgments
This work is accomplished  with a financial support of Russian
Foundation of Basic Research (Grant ü 97-01-00745).


\newpage
\begin{figure}
\caption{The integration contour used in Eq.~(29) to sum the roots
of frequency equation (20).}
\vspace{1cm}
\caption{The interquark potential $V(\rho)/M$ generated by the
modified rigid string with the Gaussian curvature in the action
at different values of the parameters
$\alpha$ and $\beta$. In Fig. $2a$ the dashed curves represent the
limiting  cases $\alpha=0$ (the Nambu--Goto string) and
$\alpha=\infty$. All the potential curves with finite values of
$\alpha$ lie between them. In Fig. $2b,c,d$
the potential for $\beta=0$ is plotted by dashed lines.
With increasing $\beta$ the potential curves are shifted to the
right  with respect to the curve corresponding to $\beta=0$.}
\end{figure}
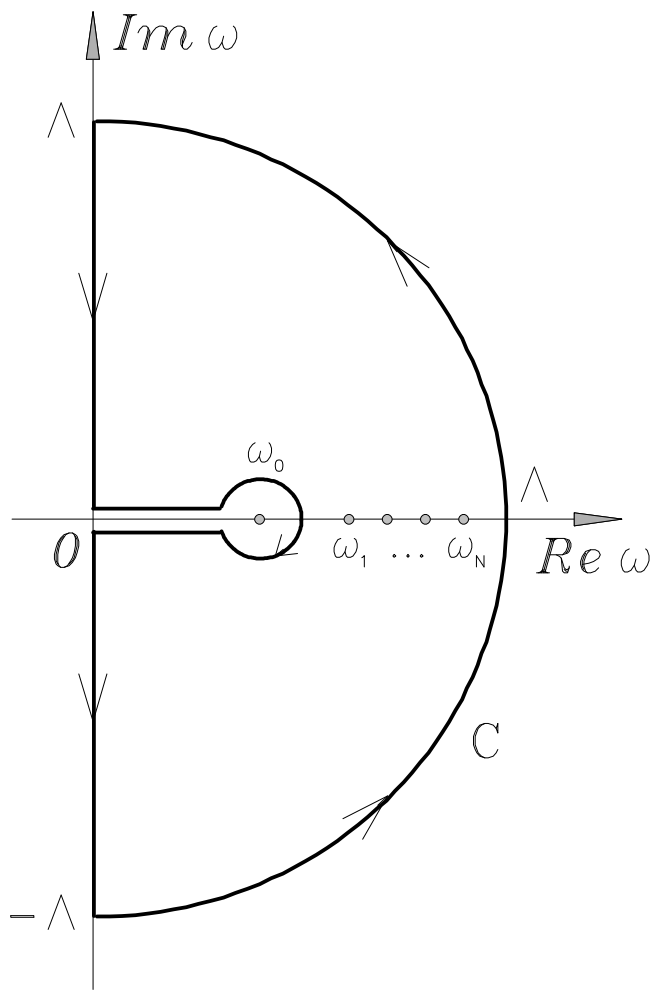
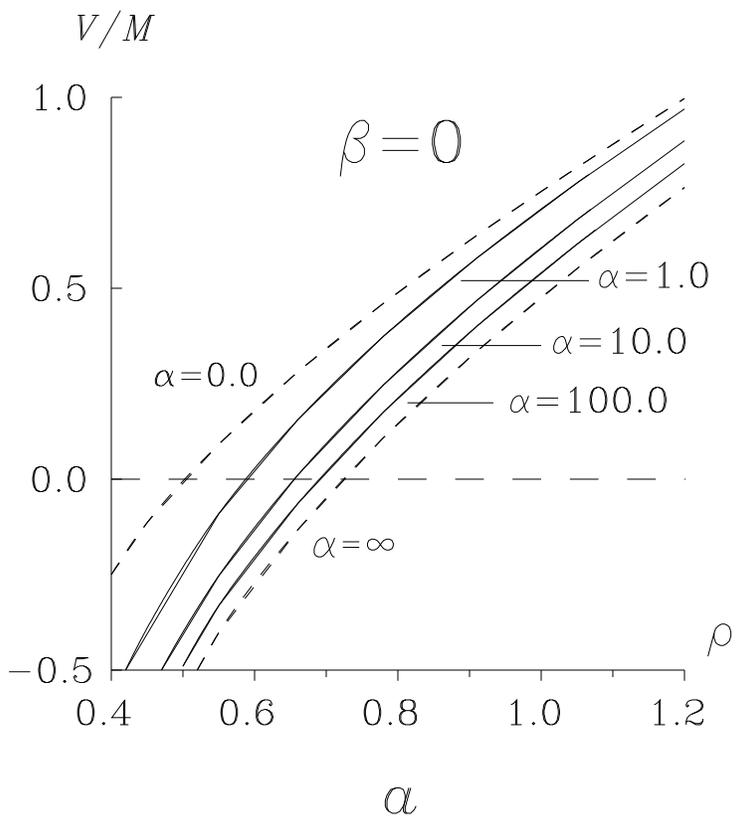
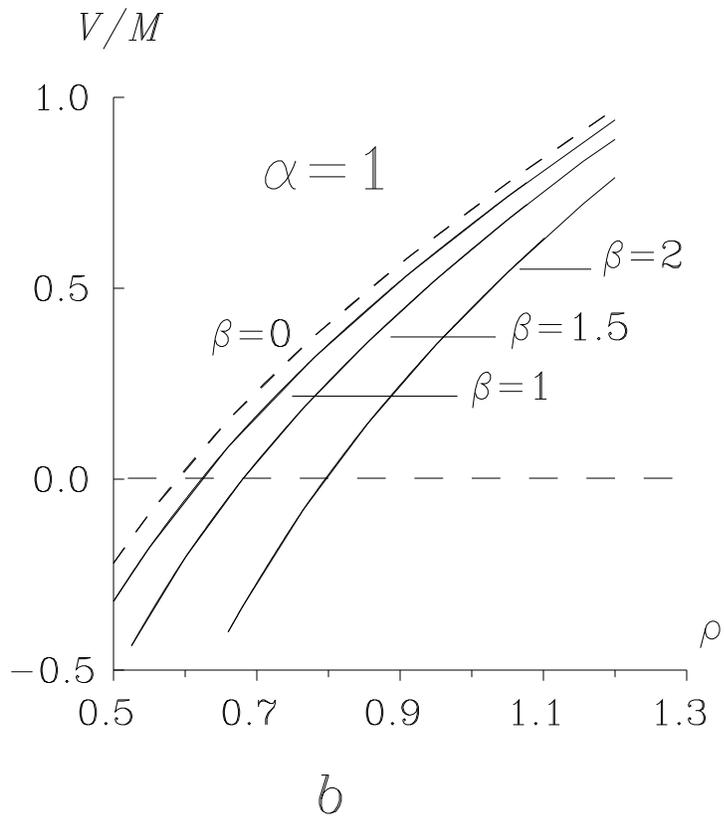
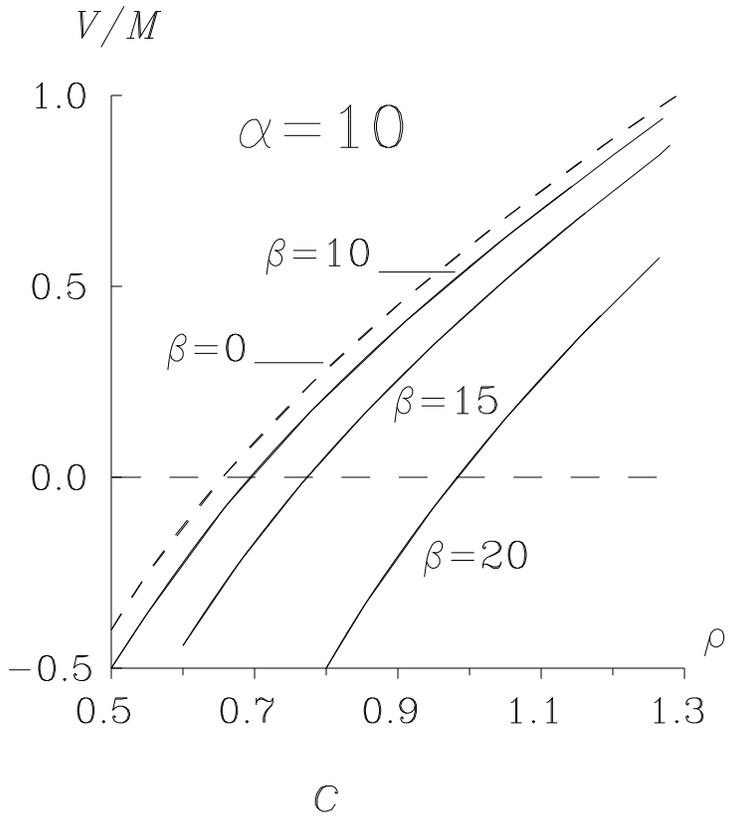
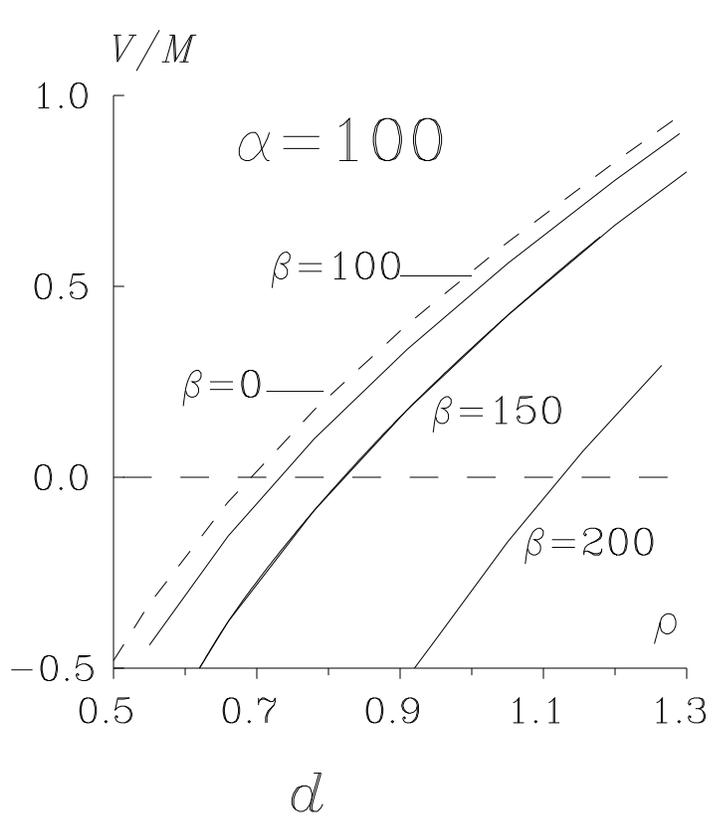
\end{document}